\documentclass[10pt,aps,prx,twocolumn,notitlepage,showpacs,superscriptaddress, 
]{revtex4-2}
\usepackage{blindtext}
\usepackage{epsfig}
\usepackage{color}
\usepackage{amsmath}
\usepackage{amsfonts}
\usepackage{amssymb}
\usepackage{graphicx}%
\usepackage{url}
\usepackage[breaklinks]{hyperref}
\usepackage[english]{babel}
\usepackage{esvect}
\usepackage[normalem]{ulem}
\usepackage[inline]{enumitem}
\usepackage{xfrac}
\usepackage{dsfont}
\usepackage{bm}
\usepackage{titlesec}
\usepackage{isomath}
\DeclareMathAlphabet\mathbfcal{OMS}{cmsy}{b}{n}

\titleformat{\section}[runin]{\normalfont \itshape}{}{}{}[\hspace{8pt}{--}]

\hypersetup{colorlinks,citecolor=blue,filecolor=blue,linkcolor=blue,urlcolor=blue}

\begin{document}

\title{Global and Local Topological Quantized Responses from Geometry, Light and Time}
\author{Karyn Le Hur}
\affiliation{CPHT, CNRS, Institut Polytechnique de Paris, Route de Saclay, 91128 Palaiseau, France}

\begin{abstract}
To describe a spin-$\frac{1}{2}$ particle on the Bloch sphere with a radial magnetic field and topological states of matter from the reciprocal space, we introduce $C$ square ($C^2$) as a local formulation of the global topological invariant. For the Haldane model on the honeycomb lattice, this $C^2$ can be measured from the Dirac points through circularly polarized light related to the high-symmetry $M$ point(s). For the quantum spin Hall effect and the Kane-Mele model, the $\mathbb{Z}_2$ topological number robust to interactions can be measured locally from a correspondence between the pfaffian and light. We address a relation with a spin pump and the quantum spin Hall conductance. The analogy between light and magnetic nuclear resonance may be applied for imaging, among other applications.

\end{abstract}
\maketitle

Topological states of matter find various applications in physics and quantum transport due to their protected edge modes and surface states \cite{QHE,RMPReview,QiZhang} which are related to the bulk of the system via a topological quantized invariant $C$ \cite{Thouless}. Circularly polarized light represents a powerful tool to detect topological properties of band structures \cite{Berkeley,Berkeley2,JarilloHerrero,Nathan,Hamburg,LightStoch}.  This topological number can be measured from the photo-currents integrated in the whole Brillouin zone and circular dichroism \cite{Nathan,Hamburg} with a correspondence to
the conductivity \cite{Thouless}. Here, we elaborate on the local definition of the global invariant $C^2$ from the geometry. We show its relevance for spin-$\frac{1}{2}$ particles and topological states of matter related to the quantum anomalous \cite{Haldane,QiZhang} and to the quantum spin Hall effects \cite{KM,BernevigZhang,Wurzburg} on the honeycomb lattice at half-filling \cite{RMPgraphene}.  For the Haldane model \cite{Haldane}, circularly polarized light can equally measure $C^2$ from the time evolution of the inter-band transition probability resolved locally in the Brillouin zone. The conductivity is also revealed from the Berry curvatures at the Dirac points. Through a protocole analogous to the nuclear magnetic resonance, we also show how light can detect the $\mathbb{Z}_2$ topological Chern number locally \cite{Sheng,KM} for the quantum spin Hall effect when the conductivity is zero. The formalism describes the topological states from the geometry starting with a radial magnetic field on the Riemann, Poincar\' e, Bloch $S^2$ sphere.

\begin{figure}[t]
\centering
\includegraphics[width=0.5\textwidth]{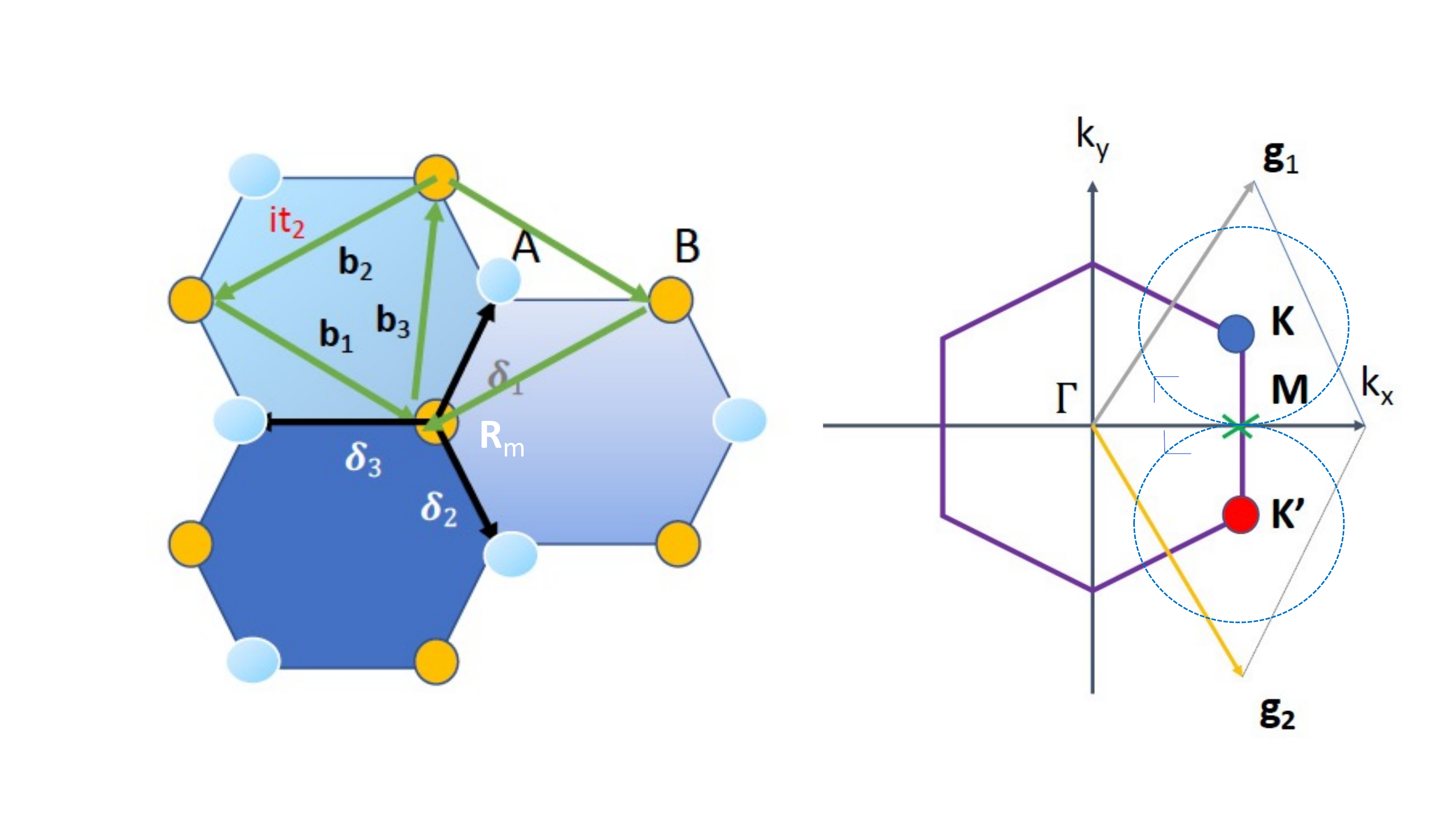}
\vskip -0.5cm
\caption{Honeycomb lattice with sub-lattices $A$ and $B$ forming the spin-$\frac{1}{2}$ (or dipole) Hilbert space. The topological Haldane model arises when including the hopping term $t_2 e^{i\phi}$ on a link through the ${\bf b}_i$ vectors with here $\phi=\frac{\pi}{2}$. Brillouin zone defined through the vectors ${\bf g}_1$ and ${\bf g}_2$. The circularly polarized lights from the $K$ and $K'$ Dirac points form a linearly polarized wave at the high-symmetry $M$ point.}
\label{topofigure}
\end{figure}

In the reciprocal space, we introduce lattice models described through the Hamiltonian ${\cal H}=\sum_{\bf k} {\cal H}({\bf k})$ and ${\cal H}({\bf k})=-{\bf d}({\bf k})\cdot {\mathbfit{\sigma}}$.  Here, the spin-$\frac{1}{2}$ is built from the $2\times 2$ Pauli matrices such that 
${\mathbfit{\sigma}}=(\sigma_x,\sigma_z,\sigma_z)$. The ${\bf d}$ vector, written as $(d_x,d_y,d_z)$ in the cartesian basis, corresponds to a radial magnetic field in the parameter space associated to the Bloch sphere of quantum spins $\frac{1}{2}$:
\begin{equation}
\label{fields}
{\bf d}({\bf k})={\bf d}(\varphi,\theta)=d(\cos\varphi\sin\theta,\sin\varphi\sin\theta,\cos\theta),
\end{equation}
where $d=|{\bf d}|$, $\theta$ is the polar angle and $\varphi$ the azimuthal angle in spherical coordinates. One important class of topological models is associated to the Haldane model on the honeycomb lattice \cite{Haldane}. The Hamiltonian here acts on the Hilbert space $\{|a\rangle;|b\rangle\}$ formed with the two sublattices $A$ and $B$ of the honeycomb lattice (see Fig. \ref{topofigure}), which allows an analogy with the spin-$\frac{1}{2}$ and a dipole \cite{RMPgraphene}. The two inequivalent Dirac points $K$ and $K'$ in the Brillouin zone correspond to the north and south poles respectively traducing the mass inversion $\pm m$ or inversion of the direction of the magnetic field at these special points \cite{SpheresArticle}. In the Haldane model, evaluating the second-nearest neighbors' hopping terms in Fig. \ref{topofigure} around $K$ and $K'$, the mass is equal to $m=d=3\sqrt{3}t_2$ where $t_2e^{i \phi}$ refers to the second nearest-neighbour hopping term and $\phi=\pi/2$ corresponds to a Peierls phase \cite{Haldane}. 

For a wave-vector ${\bf k}=f(\varphi,\theta)$, the eigenstates can be written similarly as the spin-$\frac{1}{2}$
\begin{equation}
\label{eigenstates}
|\psi_+\rangle =
\left(
\begin{array}{lcl}
\cos\frac{\theta}{2}e^{-i\frac{\varphi}{2}} \\
\sin\frac{\theta}{2} e^{i\frac{\varphi}{2}}
\end{array}
\right),
\hskip 0.25cm
|\psi_-\rangle =
\left(
\begin{array}{lcl}
-\sin\frac{\theta}{2} e^{-i\frac{\varphi}{2}} \\
\cos\frac{\theta}{2} e^{i\frac{\varphi}{2}}
\end{array}
\right).
\end{equation}
The topological Chern number, defined globally from the Brillouin zone,  on the sphere $S^2$ reads \cite{SpheresArticle}
\begin{equation}
\label{Cnumber}
C =  \frac{1}{2\pi} \int_0^{2\pi}\int_0^{\pi} F_{\varphi\theta} d\varphi d\theta,
\end{equation}
with the Berry curvature ${\bf F}=\bm{\nabla}\times{\mathbfcal{A}}$ \cite{Berry} such that $F_{\varphi\theta}=(\partial_{\varphi} {\cal A}_{\theta}-\partial_{\theta}{\cal A}_{\varphi})$. The Berry connection ${\mathbfcal{A}}=i\langle \psi| {\bf \nabla} |\psi\rangle$ plays a similar role as the vector potential in electromagnetism and momentum in quantum mechanics. For the lower energy eigenstate $|\psi_+\rangle$ corresponding to the occupied band in the honeycomb lattice model at half-filling, ${\cal A}_{\varphi}(\theta)=\frac{\cos\theta}{2}$, $F_{\varphi\theta}(\theta)=\frac{\sin\theta}{2}$ and $C=1$. For a spin-$\frac{1}{2}$, $C$ is a $\mathbb{Z}$ number equal to $0,\pm 1$ in agreement with the Poincar\' e-Hopf theorem. To derive locally the topological responses of the system, we introduce smooth fields.

These smooth fields ${\mathbfcal{A}}'$ can be built from the analogy to electromagnetism \cite{Yang}.
 The sphere with $C=1$ can be seen, from Stokes' theorem, as two regions (hemispheres) linked through an interface (boundary) corresponding to the polar angle $\theta_c$. The smooth fields on the north $(\theta<\theta_c)$ and south $(\theta>\theta_c)$ hemispheres take the precise forms ${\cal A}'_{\varphi}(\theta<\theta_c)={\cal A}_{\varphi}(\theta)-{\cal A}_{\varphi}(0)=-\sin^2\frac{\theta}{2}$ and ${\cal A}'_{\varphi}(\theta>\theta_c)={\cal A}_{\varphi}(\theta)-{\cal A}_{\varphi}(\pi)=\cos^2\frac{\theta}{2}$ \cite{SpheresArticle}.  The fields ${\cal A}_{\varphi}(0)=\frac{1}{2}$ and ${\cal A}_{\varphi}(\pi)=-\frac{1}{2}$ are uniquely defined at the poles of the sphere and importantly they are stable towards smooth deformations of the sphere as a cylinder or ellipse, such that the global topological information can be transported at the poles and at any angle $\theta_c$:
  \begin{equation}
 \label{stokes}
 C={\cal A}_{\varphi}(0)-{\cal A}_{\varphi}(\pi)={\cal A}'_{\varphi}(\theta>\theta_c)-{\cal A}'_{\varphi}(\theta<\theta_c).
 \end{equation}
 This formula is also applicable in the case of entangled spheres which will develop fractional topology \cite{SpheresArticle}. The topological number can be viewed as a charge or monopole induced by the magnetic field which produces a discontinuity of ${\cal A}'_{\varphi}(\theta)$ at $\theta=\theta_c$, such that the sphere turns into a donut or a cup. Eq. (\ref{stokes}) leads to
 \begin{equation}
 \label{C2}
 C^2 = {\cal A}'^2_{\varphi}(\theta>\theta_c) + {\cal A}'^2_{\varphi}(\theta<\theta_c) -2{\cal A}'_{\varphi}(\theta>\theta_c){\cal A}'_{\varphi}(\theta<\theta_c).
 \end{equation}

{\it Topological Responses from the spin-$\frac{1}{2}$.---} Here, we derive useful correspondences between global and local topological properties from the poles of the sphere and the Dirac cones of the honeycomb lattice. 

At the $K$ point of the Brillouin zone in Fig. \ref{topofigure}, the tight-binding model gives rise to the Dirac Hamiltonian ${\cal H}(K)=v_F(p_x\sigma_x+p_y\sigma_y)$. Close to the $K'$ point, similarly ${\cal H}(K')=v_F(p_x\sigma_x-p_y\sigma_y)$ where $v_F=\frac{3 ta}{2}$ is the Fermi velocity of graphene with $t$ the nearest-neighbour hopping amplitude and $a$ the lattice spacing. Here, ${\bf p}$ refers to a small wave-vector deviation from a Dirac point, ${\bf k}={\bf K}+{\bf p}$ and similarly for the ${\bf K'}$ point. The description around the Dirac points
here assumes values of $0<t_2<0.2t$ with a sufficiently large density of states around these points \cite{Pythtb}. Hereafter, we will show that the informations at the Dirac points are in fact related to the high-symmetry $M$ point from the lattice.

We introduce the angle $\tilde{\varphi}$ such that $p_x+i\zeta p_y = |{\bf p}| e^{i\zeta\tilde{\varphi}}$. The azimuthal angle $\varphi$ on the Bloch sphere is now related to the polar angle $\tilde{\varphi}$ associated to the cone geometry around a Dirac point. In the Haldane model, for 
$\theta\rightarrow 0$ we identify $-(d_x,d_y,d_z)=-d(\cos\varphi\sin\theta,\sin\varphi\sin\theta,\cos\theta) = (v_F|{\bf p}|\cos\tilde{\varphi}, v_F|{\bf p}|\sin\tilde{\varphi},-m)$ such that $\tilde{\varphi}=\varphi\pm\pi$ and $\tan\theta=\frac{v_F|{\bf p}|}{m}$. Around the south pole, we can modify $\varphi\rightarrow -\varphi$ and $m\rightarrow -m$ when $\theta\rightarrow \pi$ corresponding to the $K'$ point. The two Dirac cones are now centered around the two poles with a radius related to $v_F|{\bf p}|$. Related to the spin-$\frac{1}{2}$ particle, we have the identities $\frac{\partial{\cal H}}{\partial p_x} = v_F\sigma_x$ and $\frac{\partial{\cal H}}{\partial (\zeta p_y)} = v_F\sigma_y$, with 
$\zeta=\pm 1$ at the K and K' points. 

Swapping from spherical to cartesian coordinates, now we evaluate the Berry curvature \cite{Berry,Thouless,Niu} on the lattice 
\begin{equation}
\label{F0}
F_{p_x p_y}(\theta) = i\frac{(\langle \psi_-|\partial_{p_x}{\cal H}|\psi_+\rangle\langle \psi_+|\partial_{p_y}{\cal H}|\psi_-\rangle-(p_x\leftrightarrow p_y))}{(E_--E_+)^2},
\end{equation}
with $\partial_{p_x}=\frac{\partial}{\partial p_x}$ and $\partial_{p_y}=\frac{\partial}{\partial p_y}$. Here, $E_+=-d$ and $E_-=+d$ are the energies of the lower and upper bands related to $|\psi_+\rangle$ and $|\psi_-\rangle$.  From the correspondence between eigenstates in the lattice model and those of the sphere in Eqs. (\ref{eigenstates}), for $\theta\rightarrow 0$ approaching the $K$ point 
\begin{equation}
F_{p_x p_y}(\theta) = \frac{v_F^2}{2 d^2} \cos\theta.
\end{equation}
Close to $K'$ on the lattice, we have $F_{p_x -p_y}(\theta+\pi)=\frac{v_F^2}{2 d^2} \cos(\theta+\pi)=-F_{p_x p_y}(\theta+\pi)$ when $\theta+\pi\rightarrow \pi$. These relations result in the identity:
\begin{eqnarray}
\label{F}
\frac{m^2}{v_F^2}\left(F_{p_x p_y}(0) \pm F_{p_x \pm p_y}(\pi)\right) = {\cal A}_{\varphi}(0)-{\cal A}_{\varphi}(\pi) = C.
\end{eqnarray}
This implies that the quantum Hall conductivity \cite{Thouless,Niu} related to ${\cal A}_{\varphi}(0)-{\cal A}_{\varphi}(\pi)$ on the sphere \cite{SpheresArticle} is also defined from the Berry curvatures at the two Dirac points on the lattice through the identification $F_{p_x p_y}(0)=F_{p_x p_y}(K)$ and $F_{p_x p_y}(\pi)=F_{p_x p_y}(K')$. The local information encoded in the Berry fields at the Dirac points on the lattice is accessible in ultra-cold atoms \cite{MunichBerry1}. Eq. (\ref{F}) also implies that the quantity $F_{p_x p_y}(0) \pm F_{p_x \pm p_y}(\pi)$ can be directly measured locally from the photo-induced currents when coupling to circularly polarized light \cite{Nathan,Hamburg,LightStoch}.

Here, we remind that from the Ehrenfest theorem, we can also evaluate the pseudo-spin averaged magnetization $\langle \sigma_z(\theta)\rangle = \langle \psi_+ | \sigma_z |\psi_+ \rangle=\cos\theta=2{\cal A}_{\varphi}$ such that $C$ can be measured when driving from north to south pole in time since $\langle \sigma_z(0)\rangle - \langle \sigma_z(\pi)\rangle = -\int_0^{\frac{\pi}{v}} \frac{\partial\langle \sigma_z(t)\rangle}{\partial t} dt$ with the angle $\theta=vt$ \cite{SantaBarbara,Boulder,Spherebath}. This also leads to another local interpretation of $C^2$ for a spin-$\frac{1}{2}$
\begin{equation}
\label{magnetization}
2C^2 - 1 = -\langle \sigma_z(0)\rangle \langle \sigma_z(\pi) \rangle.
\end{equation}
As we show below, the quantity $C^2$ can be measured locally from the inter-band transition probabilities both for a spin-$\frac{1}{2}$ and for the topological lattice model. This quantity is in fact related to local topological marker in terms of the Berry connections that we introduce as ${\cal I}(\theta)$:
\begin{widetext}
\begin{eqnarray}
\label{Itheta}
{\cal I}(\theta) = \left\langle \psi_+ \left|\frac{\partial {\cal H}}{\partial p_x} \right|\psi_-\right\rangle \left\langle \psi_- \left|\frac{\partial {\cal H}}{\partial p_x} \right|\psi_+\right\rangle +  \left\langle \psi_+ \left|\frac{\partial {\cal H}}{\partial p_y} \right|\psi_-\right\rangle \left\langle \psi_- \left|\frac{\partial {\cal H}}{\partial p_y} \right|\psi_+\right\rangle = 2v_F^2 \left(\cos^4\frac{\theta}{2}+\sin^4\frac{\theta}{2}\right).
\end{eqnarray}
\end{widetext}
For a spin-$\frac{1}{2}$ particle, we can identify $\sigma_x=\frac{1}{v_F}\frac{\partial{\cal H}}{\partial p_x} $ and $\sigma_y=\frac{1}{v_F}\frac{\partial{\cal H}}{\partial p_y}$ related to inter-band `dipole' transitions.

This equality takes an identical form at both poles since $\zeta^2=+1$ and for one sphere ${\cal I}(0)={\cal I}(\pi)$. From Eq. (\ref{C2}) and the geometry, this results in
\begin{eqnarray}
\label{I}
\frac{{\cal I}(\theta)}{2v_F^2} = \left(2{\cal A}'_{\varphi}(\theta<\theta_c) {\cal A}'_{\varphi}(\theta>\theta_c)+C^2\right).
\end{eqnarray}
Close to the poles, we have the following relation with the square of the energetics $2{\cal A}'_{\varphi}(\theta<\theta_c) {\cal A}'_{\varphi}(\theta>\theta_c)=-\frac{v_F^2|{\bf p}|^2}{2m^2}$. 
At the two poles, we obtain
\begin{equation}
\label{equationC}
\frac{{\cal I}(0)+{\cal I}(\pi)}{4v_F^2}=C^2.
\end{equation}
 Now, we show that ${\cal I}(\theta)$ is precisely measured when coupling to circularly polarized light.

{\it Circularly Polarized Light and Time.---}  We define the vector potential ${\bf A}$ associated to the light field such that $A_x=A_0\cos\omega t$ and $A_y=\mp A_0\sin\omega t$ with $\pm$ for the right-handed $(+)$ and left-handed $(-)$ polarizations respectively according to the Jones representation of vectors. The light-matter coupling induces a dipole-light Hamiltonian $\delta {\cal H}_{\pm} = A_0 e^{\pm i\omega t} |a\rangle\langle b| +h.c.$ with 
$\sigma^+=|a\rangle\langle b|$ and $\sigma^-=|b\rangle \langle a|$ producing inter-band transitions \cite{LightStoch}. The resonance situation is obtained from the transformation $|b\rangle=e^{\mp i \frac{\omega t}{2}}|b'\rangle$ and $|a\rangle=e^{\pm i \frac{\omega t}{2}}|a'\rangle$ such that $E_b-E_a=\pm \hbar\omega$ for the $(\pm)$ polarization. Via the Fermi golden rule, we obtain the inter-band transition rates:
\begin{equation}
\label{equations}
\Gamma_{\pm}(\omega) = \frac{2\pi}{\hbar}\left| \langle \psi_- | \delta {\cal H}_{\pm} |\psi_+\rangle \right|^2 \delta(E_b - E_a \mp \hbar\omega).
\end{equation}
Around the $K$ point, we have $E_b-E_a=E_- - E_+=2m=\hbar\omega$ and around the $K'$ point we have $E_b-E_a=E_+ -E_-=-2m=-\hbar\omega$. For frequencies $\omega>0$, one light polarization resonates with one Dirac point as long as we are in the topological phase. 
For $\theta\rightarrow 0$ and $\pi$, 
\begin{equation}
\label{Gamma}
\Gamma_{\pm}(\theta,\omega) = \frac{2\pi}{\hbar} A_0^2 \left(\frac{{\cal I}(\theta)}{2 v_F^2}\right) \delta(E_b -E_a \mp \hbar\omega),
\end{equation}
with $\hbar=\frac{h}{2\pi}$ the Planck constant. Here, we underline that Eq. (\ref{Gamma}) is also valid at the high-symmetry $M$ point between $K$ and $K'$ from the properties of the lattice model only \cite{SM,FuKane} using the fact that ${\cal I}(\theta)$ is invariant under $k_y\rightarrow -k_y$ or $\varphi\rightarrow -\varphi$. This is equivalent to define
 \begin{equation}
 \label{M}
{\cal I}\left(\frac{\pi}{2}\right)={\cal I}(M)=\frac{C^2}{2}(2v_F^2).
 \end{equation}
At the $M$ point, each light polarization contributes to a prefactor $\frac{C^2}{2}$ and the superposition of the two light polarizations is equivalent to a linearly polarized wave along 
$x$ direction (see Fig. \ref{topofigure}).

It is now judicious to introduce the frequency-integrated rates $\int_0^{+\infty} \Gamma_{\pm}(\theta,\omega)d\omega=\frac{1}{2}\int_{-\infty}^{+\infty} \Gamma_{\pm}(\theta,\omega)d\omega=\frac{{\Gamma}_{\pm}(\theta)}{2}$ with the identifications $\theta=0=K$ and $\theta=\pi=K'$. The interesting observation here is that the local quantity
\begin{equation}
\Delta \Gamma = \frac{1}{2\pi}\left(\frac{\Gamma_+(K) + \Gamma_-(K')}{2}\right) = \frac{1}{\hbar^2}A_0^2 C^2,
\end{equation}
is measurable from circularly polarized light. The factor $C^2$ in the transition rates, defined locally from Eq. (\ref{equationC}), was not precisely identified in the literature previously.
This formula can find applications for driven spin models where ${\bf A}$ represents a rotating magnetic field in the $xy$ plane as in nuclear magnetic resonance (NMR). 

This result can be verified when calculating the inter-band transition probability or spin-flip probability ${\cal P}(\tilde{\omega},t)=|\langle \psi_-|\psi_+(t)\rangle|^2$ in real time at the $K$ or $K'$ Dirac point, with here $\tilde{\omega}=\omega-2m$.  Preparing the system at time $t=0$ in $|\psi_+\rangle$, we find \cite{SM}
\begin{equation}
\label{transition}
{\cal P}(\tilde{\omega},t)=\frac{4A_0^2}{(\hbar\tilde{\omega})^2}C^2\sin^2\left(\frac{1}{2}\tilde{\omega}t\right).
\end{equation}
The evolution of the (normalized) lowest-band population in real time then satisfies accordingly $N_+(t) = |\langle \psi_+(t) |\psi_+(t)\rangle|^2= N_+(0) - {\cal P}(\tilde{\omega},t) = 1-{\cal P}(\tilde{\omega},t)$. This mediates inter-band transitions, in agreement with the Rabi formula for NMR and applications to quantum Hall systems \cite{Halllight}, where we observe an additional topological prefactor coming from the effect of the radial magnetic field for topological Bloch bands. 
If we select the light frequency at resonance $\tilde{\omega}=\omega-\frac{2m}{\hbar} \rightarrow 0$, ${\cal P}(\tilde{\omega},t)=\Delta\Gamma t^2$ when $\Gamma_{\pm}(\theta,\omega)$ are evaluated at $\theta\rightarrow 0$ and $\theta\rightarrow \pi$. When sweeping on light frequencies, we find $\frac{d N_+}{dt} = -\pi\Delta\Gamma$.  In the Supplemental Information \cite{SM}, we justify the relation between Eq. (\ref{transition}) with the photo-currents and discuss the topological protection towards a Semenoff mass. We also emphasize here that the light response at the $M$ point would allow to directly measure a $\frac{1}{2}$-response for topological semi-metallic bilayers from one light polarization only related to the topological Dirac point \cite{SpheresArticle}. These results related to inter-band transition probabilities in time are observable with current technology as in ultra-cold atoms \cite{Munich}. Related to possible applications of circularly polarized light in imaging, the Fourier transform of the signal gives rise to resonant $\delta$-peaks.  
 
{\it Quantum Spin Hall Effect, $\mathbb{Z}_2$ topological number from Light.---} Here, we generalize the analysis to two spheres described by the two Hamiltonians ${\cal H}_1({\bf k})=-{\bf d}_1({\bf k})\cdot\mathbfit{\sigma}_1$  and ${\cal H}_2({\bf k})={\bf d}_2({\bf k})\cdot\mathbfit{\sigma}_2$ with $d_{1x}=d_{2x}$, $d_{1y}=d_{2y}$, $d_{1z}=m_1$ and $d_{2z}=m_2$. The situation with $m_1=m_2=m$ finds applications in the Kane-Mele model on the honeycomb lattice \cite{KM} with spin-orbit coupling where $1,2$ correspond to the two spin-polarizations of a spin-$\frac{1}{2}$. Asymmetric masses $m_1\neq m_2$ can occur in a bilayer structure \cite{bilayermodel}. For each sphere $\alpha=1,2$, the pseudospin-$\frac{1}{2}$ is built from the Pauli matrices acting on the Hilbert space $\{ a_{\alpha};b_{\alpha} \}$ associated to the occupancy on a sub-lattice of the honeycomb lattice. Going from sphere $1$ to $2$ is equivalent to change the role of the lower and upper energy eigenstates in Eq. (\ref{eigenstates}) and to adjust the topological numbers as $C_1=+1$ and $C_2=-1$.  This modifies ${\cal A}'_{\varphi}(\theta>\theta_c)\rightarrow - {\cal A}'_{\varphi}(\theta>\theta_c)=-\cos^2\frac{\theta}{2}$ and ${\cal A}'_{\varphi}(\theta<\theta_c)\rightarrow -{\cal A}'_{\varphi}(\theta<\theta_c)=\sin^2\frac{\theta}{2}$ translating the $\mathbb{Z}_2$ symmetry. 
Generalizing the definition of Eq. (\ref{Itheta}) for each sphere, the light measures the spin Chern number $C_s=\sum_{\alpha} C_{\alpha}^2=C_1-C_2=+2$ \cite{Sheng} locally from the Dirac points whereas the quantum Hall conductivity measures $\sum_{\alpha} C_{\alpha}=0$. We obtain additional information on the detuning effects from the analogy with the NMR  \cite{SM}. 

{\it Pfaffian, topological $\mathbb{Z}_2$ number and Light Response.---} To establish a correspondence with the Pfaffian for the Kane-Mele model \cite{KM2}, we can simply identify the two lowest filled energy bands on the lattice with eigenstates $|u_i({\bf k})\rangle$ for $i=1,2$ to the Bloch sphere description. We have $|u_1({\bf k})\rangle=|\psi_+(\theta,\varphi)\rangle$ and $|u_2({\bf k})\rangle = |\psi_-(\theta,\varphi)\rangle$. Time-reversal symmetry $\Theta$ modifies the spin magnetization $\tau_z\rightarrow -\tau_z$, defined as $\tau_z=\frac{1}{2}(\sigma_{1z}-\sigma_{2z})$, and ${\bf k}\rightarrow -{\bf k}$ in the Hamiltonian such that $\Theta {\cal H}_1({\bf k})\Theta^{-1}= {\cal H}_2(-{\bf k})$ and $\Theta {\cal H}_2({\bf k}) \Theta^{-1}= {\cal H}_1(-{\bf k})$. We can evaluate the Pfaffian $Pf_{ij} = \epsilon_{ij} \langle u_i({\bf k}) | \Theta | u_j({\bf k})\rangle$  on the sphere identifying $\Theta|u_i({\bf k})\rangle = \epsilon_{ij}|u_j(-{\bf k})\rangle^*$. Within our definition of the Brillouin zone, the transformation ${\bf k}\rightarrow -{\bf k}$ is equivalent to $k_y\rightarrow -k_y$ such that $Pf_{12}=\langle \psi_+(0)|\psi_+(\pi)\rangle^*$.  
The zeros of the Pfaffian at the poles of the sphere are then related to the perfect quantization of the light response 
\begin{equation}
\frac{1}{2v_F^2}\left({\cal I}_1(\theta)+{\cal I}_2(\theta)\right) = C_s -P({\bf k})^2.
\end{equation}
Here, ${\cal I}_i(\theta)$ is generalized from Eq. (\ref{I}) for each spin polarization and we identify $Pf_{12}=Pf_{21}=P({\bf k})=\sin\theta$. We also verify the equivalent form $P({\bf k})=\frac{v_F|{\bf p}|}{m}$ close to the Dirac points from the eigenstates on the lattice.
Measuring the light responses at the Dirac points corresponds to detect the $\mathbb{Z}_2$ spin Chern number from the zeros of the Pfaffian.
 
{\it Quantum Spin Pump and Interactions.---}  Here, we show that the local light response is stable towards general perturbations such as a Rashba spin-orbit interaction making a link with a $\mathbb{Z}_2$ quantum spin pump. From the discussion around Eq. (\ref{magnetization}), we can also relate the local spin magnetizations to the topological $\mathbb{Z}_2$ number: 
\begin{equation}
\label{Cs}
\langle \tau_z(0)\rangle - \langle \tau_z(\pi)\rangle = -\int_0^{\frac{\pi}{v}} \frac{\partial\langle \tau_z(t)\rangle}{\partial t}dt = C_s.
\end{equation}
Within the quantum spin Hall phase, the topological charges $C_1=+1$ and $C_2=-1$ will remain identical and similarly for the local spin magnetizations and the light responses. We can formulate this conclusion more quantitatively writing a two-spheres' wave-function $|\psi\rangle = \sum_{kl} c_{kl}(\theta) |\Phi_k\rangle_1\otimes |\Phi_l\rangle_2$ \cite{SpheresArticle} with a choice of Hilbert space related to Eq. (\ref{eigenstates}), 
$|\Phi_+\rangle = 
\left(
\begin{array}{c}
e^{-i\frac{\varphi}{2}} \\
0
\end{array} \right)$
and 
$
|\Phi_-\rangle = 
\left(
\begin{array}{c}
0 \\
e^{i\frac{\varphi}{2}} 
\end{array} \right)$. Here, $|\Phi_+\rangle$ and $|\Phi_-\rangle$ refer to projections on sub-lattice $A$ or $B$ for a spin polarization. 
The function $c_{kl}(\theta)=c_{k}^1(\theta)c_{l}^2(\theta)$ with $k,l=\pm$ is independent of $\varphi$ close to the poles because all $\varphi$ angles are equivalent. Then, this gives rise to the identities $\langle \psi| \tau_z | \psi\rangle = |c_{+-}|^2-|c_{-+}|^2={\cal A}_{\varphi}^1-{\cal A}_{\varphi}^2$. 
Introducing the gauge invariant quantities ${\cal A}_{\varphi}^i(0)-{\cal A}_{\varphi}^i(\pi)=C_i$ for $i=1,2$ we verify the validity of Eq. (\ref{Cs}) with $C_s=+2$. The robustness of $C_s$ \cite{Sheng} comes from the fact that as long as we stay therein
the topological insulator phase the coefficients $c_{--}$ and $c_{++}$ remain zero at the poles of the sphere or at the Dirac points on the lattice. The robustness of the light response is implicitly driven from Stokes' theorem in Eq. (\ref{stokes}). Including a Hubbard interaction \cite{SM} the local light responses remain quantized in a many-body sense in the topological phase(s) until quantum phase transitions such as Mott phases \cite{LightStoch,KMstoch}.

{\it Correspondence with Edge Modes on a Cylinder. ---} Here, we show the correspondence with a Laughlin cylinder geometry. The cylinder acts in the reciprocal space of the lattice model with periodic boundary conditions in $k_x$ direction. If we define the Berry curvature ${\bf F}=\frac{1}{2}{\bf e}_r$ along the
radial direction on the surface of the cylinder then we can adjust its height to $H=2$ such that the topological number reproduces $C=\frac{1}{2\pi}\int_0^H dz\int_0^{2\pi} d\varphi {\bf F}\cdot{\bf e}_r = 1$ for the Haldane model. Related to the Brillouin
zone, the $z$ variable is defined such that $H=2$ refers to the distance $\frac{4\pi}{3\sqrt{3}a}$ between $K$ and $K'$ along $k_y$ direction. The vector potential
can be defined as ${\cal A}_{\varphi}(z)=\frac{z}{2}$ such that at the north disk ${\cal A}_{\varphi}=+\frac{1}{2}$ and at the south disk ${\cal A}_{\varphi}=-\frac{1}{2}$. From the spherical coordinates $z=\cos\theta$, then ${\cal A}_{\varphi}(\theta)=\frac{\cos\theta}{2}$ producing the same smooth fields on the cylinder ${\cal A}'_{\varphi}(z>0)={\cal A}'_{\varphi}(\theta<\theta_c)$ and ${\cal A}'_{\varphi}(z<0)={\cal A}'_{\varphi}(\theta>\theta_c)$ with here $\theta_c=\frac{\pi}{2}$. For the Kane-Mele model, we have two cylinders
such that ${\bf F}_1={\bf F}$ and ${\bf F}_2=-{\bf F}$. 

To activate the spin pump we apply an electric field ${\bf E}$ parallel to the polar angle, from north to south pole on the sphere, acting on a charge $q$ such that from Newton equation $\theta(t)=vt$ with $v=\frac{q E}{\hbar}$ in Eq. (\ref{Cs}). From the Parseval-Plancherel theorem \cite{SpheresArticle}, this produces transverse currents on the two spheres related to the smooth fields $J_{\perp}^1=J_{\perp}(\theta)=\frac{e}{t}{\cal A}'_{\varphi}(\theta<\theta_c)$ and $J_{\perp}^2=-J_{\perp}(\theta)$. To relate with the light response, we navigate such that $\theta\in[0;\pi]$
in a time $T=\frac{h}{2qE}$ producing a spin current $J_{\perp}^1-J_{\perp}^2=\frac{2q^2}{h}C_s E$. The factor $2$ specifies that a charge $-q$ also navigates in opposite direction. On the cylinder, we have the same spin current from the smooth fields identification. If we introduce a voltage drop on the cylinders $EH=(V_t-V_b)$ we verify the formation of edge modes at the boundaries with the disks, $J_{\perp}^1-J_{\perp}^2=G_s(V_t-V_b)$ and $G_s=\frac{q^2}{h}C_s$.

{\it Conclusion.---}  We have introduced a local marker to the global invariant $C^2$ with direct applications for spin-$\frac{1}{2}$ particles and topological lattice models. We have also shown that the quantum Hall conductivity can be related to the Berry curvatures locally at the Dirac points on the honeycomb lattice. These predictions may find applications in quantum materials \cite{Wurzburg,Bismuth} and ultra-cold atoms \cite{MIT} related to developments in spintronics and light-induced quantized local responses. As further perspectives, we highlight here that the formalism may equally describe topological superconducting wires \cite{Alicea} and three-dimensional Weyl semi-metals \cite{Nagaosa}.

K.L.H. acknowledges discussions related to the preparation of the class PHY552 at Ecole Polytechnique and with Joshua Benabou, Ephraim Bernhardt, Frederick del Pozo, Nathan Goldman, Adolfo Grushin, Joel Hutchinson, Philipp W. Klein, Julian Legendre and Han Yu Sit.
This work was supported by the french ANR BOCA and the Deutsche Forschungsgemeinschaft (DFG), German Research Foundation under Project No. 277974659.

\onecolumngrid
\subsection{${\cal I}(\theta)$ at the $M$ point from the lattice}

Here, we show a derivation of ${\cal I}(\theta)$ at the $M$ point on the lattice related to the light response.
Correspnding to Fig. 1 in the article, if we use the Bravais lattice vectors ${\bf u}_1=-{\bf b}_2=\frac{a}{2}(3,\sqrt{3})$ and ${\bf u}_2 = {\bf b}_1 = \frac{a}{2}(3,-\sqrt{3})$, we can write the graphene Hamiltonian at the $M$ point in the form
\begin{equation}
{\cal H}(M) =  w\sigma^+ +h.c.,
\end{equation}
with 
\begin{equation}
\label{w}
w = t\left(1+\sum_{i=1}^2 e^{-i {\bf k}\cdot{\bf u}_i}\right),
\end{equation}
and $k^M_x=\frac{2\pi}{3a}$, $k^M_y=0$. 

We can justify the choice of local gauge in Eq. (\ref{w}) as follows. 
Within our definition of the Brillouin zone, at this $M$ point since $k_y=0$, the Hamiltonian should be invariant under the symmetry $k_y\rightarrow -k_y$ which implies that the term $d_2\sigma_y$ in the formulation of Fu and Kane (Ref. [27] in the Letter) should be defined to be zero. The Hamiltonian at this $M$ point should be equivalent to $d_1\sigma_x=d_1\hat{P}$, with $d_1=\frac{1}{2}\left(w+w^*\right)$ and with $\hat{P}$ defined to be the parity
operator defined in a middle of a bond in a unit cell in real space, corresponding then to interchange $A\leftrightarrow B$ sublattices through the transformation $x\rightarrow -x$ or $k_x\rightarrow -k_x$. This $M$ point in the middle of $K$ and $K'$ is in fact special from the classification of $\mathbb{Z}_2$ topological insulators since $sgn(d_1)=-1$ within our definitions of $w$ whereas at the other high symmetry points, we find $sgn(d_1)=+1$. These definitions are also in agreement with the fact that the light-matter response is invariant under $\varphi\rightarrow -\varphi$. 

This results in
\begin{eqnarray}
\frac{\partial{w}}{\partial k_x} = -i t (2u_x)sgn(d_1) = (3ita).
\end{eqnarray}
Here, $sgn(d_1)=-1$ traduces that the eigenvalue of the $\sigma_x$ or parity operator on the lattice takes a negative value at this specific point, as in the definition of the $\mathbb{Z}_2$ topological invariant formulated by Fu and Kane.
In a similar way, 
\begin{eqnarray}
\frac{\partial{w}}{\partial k_y} = 0.
\end{eqnarray}
Then, we obtain
\begin{eqnarray}
\left\langle \psi_+ \left|\frac{\partial {\cal H}}{\partial k_x} \right|\psi_-\right\rangle \left\langle \psi_- \left|\frac{\partial {\cal H}}{\partial k_x} \right|\psi_+\right\rangle = 4v_F^2\cos^4\frac{\theta}{2}.
\end{eqnarray}
Here, we take into account the $\delta(E_b-E_a\mp \hbar\omega)$ function in Eq. (14) in the article such that either $w\sigma^+$ or $w\sigma^-$ contributes for a given light polarization. 
Since the sine and cosine functions are equal at $\theta=\frac{\pi}{2}$, it allows us to verify that this quantity at the $M$ point is also equal to ${\cal I}(\theta)$. At the $M$ point, then we have
\begin{eqnarray}
\left\langle \psi_+ \left|\frac{\partial {\cal H}}{\partial k_x} \right|\psi_-\right\rangle \left\langle \psi_- \left|\frac{\partial {\cal H}}{\partial k_x} \right|\psi_+\right\rangle = {\cal I}(\theta),
\end{eqnarray}
for all values of $t_2$. We also have the correspondence
\begin{equation}
{\cal I}(M) = \frac{{\cal I}(0)}{2} = \frac{{\cal I}(\pi)}{2},
\end{equation}
in the light response. 

These equations are also in agreement with the fact that the addition of the electric fields around the two Dirac points produces an electric field along $x$ direction at the $M$ point, similarly as a linearly polarized wave:
$$
{\bf E} = {\bf E}_+ + {\bf E}_- = 2 e^{i\frac{\pi}{2}}A_0 \omega e^{-i\omega t} {\bf e}_x.
$$
For a given light polarization, from the geometry, the response of the system will be halved compared to the Dirac points because $\frac{\partial w}{\partial k_y}=0$. 

\subsection{Evolution in Time and Local Photo-Induced Currents}

To study the time dynamics around the Dirac points, we write the light-matter coupling in the eigenstates' basis and apply the evolution operator in time. We redefine the eigenstates $|\psi_+(\theta=0)\rangle = -|\psi_-(\theta=\pi)\rangle$ and $|\psi_-(0)\rangle = |\psi_+(\theta=\pi)\rangle$.
At the north pole, we can write the light-matter coupling from the right-handed $(+)$ polarization as
\begin{equation}
\label{evolve}
\delta{\cal H}_+ = A_0 e^{i\omega t} e^{-i\varphi} {\cal A}_{\varphi}'(\theta>\theta_c) |\psi_-\rangle \langle \psi_+| +h.c.
\end{equation}
and ${\cal A}'_{\varphi}(\theta>\theta_c)=C$. 
Then, developing the evolution operator in time to first order in $\delta{\cal H}_+$, we have 
\begin{equation}
 |\psi_{+}(t)\rangle = e^{\frac{i}{\hbar}m t}\left(|\psi_{+}(0)\rangle - \frac{1}{\hbar\tilde{\omega}} A_0 e^{-i\varphi} {\cal A}'_{\varphi}(\theta>\theta_c)\left(e^{i\tilde{\omega}t} -1\right)|\psi_-(0)\rangle\right),
\end{equation}
with $\tilde{\omega}=\omega-2m/\hbar$ and $|\psi_-(0)\rangle=|b\rangle$. 
Here, we have selected the right-handed polarisation term $\delta {\cal H}_+$ because in the limit $t\rightarrow +\infty$ we verify that it satisfies the energy conservation in agreement with the Fermi golden rule approach. In this way, we obtain the transition probability ${\cal P}(\tilde{\omega},t)$ to reach the upper energy band 
\begin{equation}
\label{short}
{\cal P}(\tilde{\omega},t) = |\langle b | \psi_{+}(t)\rangle|^2 = \frac{4 A_0^2}{(\hbar\tilde{\omega})^2} ({\cal A}'_{\varphi}(\theta>\theta_c))^2 \sin^2\left(\frac{1}{2}\tilde{\omega}t\right).
\end{equation}
Here, $|b\rangle$ represents the upper energy state at the north pole or the $K$ Dirac point in the topological band structure. This formula is reminiscent of the nuclear magnetic resonance inter-band transition formula where we identify an additional geometrical factor encoding the topological properties
from the radial magnetic field. At short times, we obtain
\begin{equation}
{\cal P}(dt^2) = \frac{A_0^2}{\hbar^2} ({\cal A}'_{\varphi}(\theta>\theta_c))^2 dt^2,
\end{equation}
such that
\begin{equation}
\label{density}
\frac{dN_+}{dt^2} = - \frac{A_0^2}{\hbar^2} ({\cal A}'(\theta>\theta_c))^2 = - \frac{A_0^2}{\hbar^2} C^2 = -\Delta\Gamma.
\end{equation}
Here, $N_+(t) = |\langle \psi_+(t) |\psi_+(t)\rangle|^2= N_+(0) - {\cal P}(t) = 1-{\cal P}(t)$ describes the normalized number of particles in the lowest band at time $t$. This equation shows locally the relation with the smooth fields and the global topological invariant $C^2$. In fact, if we select the resonance frequency $\tilde{\omega}\rightarrow 0$, then the relation
\begin{equation}
{\cal P}(t) \sim \frac{A_0^2}{\hbar^2} C^2 t^2,
\end{equation}
can be measured for long(er) times. 

If we integrate Eq. (\ref{short}) on the light frequencies $\tilde{\omega}$, then we verify that the response becomes linear in time and the same information on $C^2$ remains. The fonction is symmetric in $\tilde{\omega}$ and therefore using the mathematical identity
\begin{equation}
\int_{-\infty}^{+\infty} \frac{e^{i\tilde{\omega}t}}{\tilde{\omega}^2} d\tilde{\omega} = -\pi t sgn(t),
\end{equation}
which goes to zero when $t\rightarrow 0^+$, then this results in 
\begin{equation}
\int_0^{+\infty} {\cal P}(\tilde{\omega},t) d\tilde{\omega} = \frac{A_0^2}{\hbar^2}C^2 \pi t.
\end{equation}
Then, we find:
\begin{equation}
\frac{dN_+}{dt} = -\frac{\pi}{\hbar^2}A_0^2 C^2 = - \pi\Delta\Gamma.
\end{equation}

To describe the physics at the $K'$ point, we can move the interface close to $\theta_c\rightarrow \pi$, such that we have the identification $C=-{\cal A}'_{\varphi}(\theta<\theta_c)$ for the left-handed polarisation
\begin{eqnarray}
\delta{\cal H}_- &=& A_0 e^{i\omega t} e^{-i\varphi}(-C)|\psi_+(\pi)\rangle \langle \psi_-(\pi)| +h.c.  \\ \nonumber
&=& A_0 e^{i\omega t} e^{-i\varphi}C|\psi_-(0)\rangle \langle \psi_+(0)| +h.c.
\end{eqnarray}
We obtain a similar formula as in Eq. (\ref{evolve}) close to the $K$ point. 

From the time evolution of the population in the lower band (or equivalently upper band), the two light polarizations play a symmetric role one at a specific Dirac point $K$ or $K'$.

We can equally calculate the photo-induced current for the light polarization $\pm$ (see for instance Ref. [10] in the article). The currents ${\bf j}_{\pm}=\pm j_{\pm}{\bf e}_{\varphi}$ turn along the azimuthal unit vector in different directions such that:
\begin{equation}
\label{Kstructure}
j_{\pm,\zeta}(t) = \frac{1}{2\hbar}A_0e^{i\omega t}\left(\frac{\partial H}{\partial(\zeta p_y)} \mp i \frac{\partial H}{\partial p_x}\right)+h.c.
\end{equation}
We remind that $\zeta=\pm 1$ at the $K$ and $K'$ Dirac points respectively from the definitions in the Letter. 
The photocurrents produce a similar result as $\Delta\Gamma$ from the Fermi golden rule, but for the currents one must then define $|{\bf\Gamma}_+(K) - {\bf\Gamma}_-(K')|$. This response is related to $F_{p_x p_y}(K)-F_{p_x -p_y}(K')$ and gives a similar result as $\Delta\Gamma$ with $|C|$ instead of $C^2$.
We have $|C|=C^2$ for a topological state with $C=1$ such that $\frac{dN_+}{dt^2}$ and $\frac{dN_+}{dt}$ are always definite negative. The inter-band transition probabilities measured locally in the reciprocal space are then related to the induced photo-currents. 

These relations show that changing of polarization is equivalent to change $\varphi\rightarrow -\varphi$ and therefore this is also equivalent to change of Dirac point since this transformation implies that $p_y\rightarrow -p_y$. 

For the Kane-Mele model, the light responses of the two spheres will be additive. Another way to interpret this result is as follows. The $+$ light polarization will resonate with sphere $1$ at the $K$ point and with sphere $2$ at $K'$ which corresponds to change $p_y\rightarrow -p_y$ for sphere $2$ in $j_+(K')$ compared to $j_+(K)$. Similarly, the $-$ light polarization will resonate with sphere $1$ at $K'$ and with sphere $2$ at $K$ which corresponds then to modify $p_y\rightarrow -p_y$ in $j_-(K)$ for sphere $2$ compared to $j_-(K')$. This change of $p_y\rightarrow -p_y$ in the formulas for the sphere $2$ compared to sphere $1$ then gives the following structure from the Fermi golden rule $|{\bf\Gamma}^1_+(K) - {\bf\Gamma}^1_-(K')-{\bf\Gamma}^2_+(K')+{\bf\Gamma}^1_-(K)|$, related to the currents, providing another physical understanding for the occurrence of $C_1-C_2$ in the light response. 

Below, we reproduce this argument in the rotating frame discussing also detuning effects. 

\subsection{Protection Towards a Semenoff Mass}

The formalism derived at the poles of the sphere is also practical to study the effect of a Semenoff mass ${\cal M}\sigma_z$ in the Hamiltonian and show the topological protection of the results for ${\cal M}<m$. 

In the vicinity of the poles, we can yet write down eigenstates in the same form as in Eq. (2) of the article. Close to the north pole of the sphere corresponding to the $K$ point on the lattice, we identify
\begin{equation}
\cos\theta = \frac{\tilde{d}_z(0)}{\sqrt{\tilde{d}_z(0)^2 +v_F^2 |{\bf p}|^2}}
\end{equation}
with
\begin{equation}
\tilde{d}_z(0)=d_z-{\cal M}=m-{\cal M}.
\end{equation}
Close to the north pole, then we have
\begin{equation}
{\cal A}_{\varphi}(\theta,\tilde{d}_z(0))=\frac{\cos\theta}{2}.
\end{equation}
Similarly, close to the south pole of the sphere corresponding to the $K'$ point on the lattice, we identify
\begin{equation}
\cos\theta = \frac{\tilde{d}_z(\pi)}{\sqrt{\tilde{d}_z(\pi)^2 +v_F^2 |{\bf p}|^2}}
\end{equation}
with
\begin{equation}
\tilde{d}_z(\pi)=-d_z-{\cal M}=-m-{\cal M}.
\end{equation}
In this way, close to the south pole, we can yet write down
\begin{equation}
{\cal A}_{\varphi}(\theta,\tilde{d}_z(\pi))=\frac{\cos\theta}{2}.
\end{equation}
Precisely at the poles we have $\sin\theta\rightarrow 0$ corresponding to $v_F|{\bf p}|\rightarrow 0$. Then, as long as ${\cal M}<m$ meaning that we are in the same topological phase, we verify that ${\cal A}_{\varphi}(0)-{\cal A}_{\varphi}(\pi)=C=+1$ with ${\cal A}_{\varphi}(0)=+\frac{1}{2}$ and ${\cal A}_{\varphi}(\pi)=-\frac{1}{2}$. 

Close to the north pole, this ensures that the smooth field ${\cal A}'_{\varphi}(\theta>\theta_c)$ keeps the same form $\cos^2\frac{\theta}{2}$. Similarly, close to the south pole, this ensures that the smooth field ${\cal A}'_{\varphi}(\theta<\theta_c)$ keeps the same form $-\sin^2\frac{\theta}{2}$. Following these facts, the results derived close to the poles for $\Gamma_+(\theta\rightarrow 0,\omega)$  and $\Gamma_+(\theta\rightarrow \pi,\omega)$ remain identical and similarly for the time responses derived below assuming that we adjust the resonance frequency accordingly such that $\hbar\omega=2\tilde{d}_z(0)$ at the north pole for the $(+)$ light polarization and $\hbar\omega=-2\tilde{d}_z(\pi)$ for the $(-)$ light polarization at the south pole. Eq. (14) in the article is also valid at the $M$ point showing that from Stokes' theorem we can teleport the information from the
poles to the equatorial plane. Since the light field is equivalent to a vector oriented along the $x$ direction, this implies that for a given light polarization, the response at the $M$ point should be halved compared to the points $K$ and $K'$, as also mentioned in Sec. A above. This argument emphasizes that ${\cal I}(M)=\frac{{\cal I}(0)}{2}=\frac{{\cal I}(\pi)}{2}$.

The topological transition at $m={\cal M}$ corresponds to ${\cal A}_{\varphi}(\theta=0)=0$ taking the limit $v_F|{\bf p}|=0^+$ and ${\cal A}_{\varphi}(\theta=\pi)=-\frac{1}{2}$. Above the transition ${\cal M}>m$, the two Dirac points would respond to one light polarization only. 

For completness, for ${\cal M}<m$, Eq. (8) in the article now takes the form
\begin{equation}
\frac{\tilde{d}_z^2(0)}{v_F^2}F_{p_x p_y} \mp \frac{\tilde{d}_z^2(\pi)}{v_F^2}F_{p_x\mp p_y} = {\cal A}_{\varphi}(0) - {\cal A}_{\varphi}(\pi)=C.
\end{equation}

\subsection{Light Responses in the Kane-Mele model}

Here, we study the light responses in the Kane-Mele model at the Dirac points from the rotating frame.
In the vicinity of the K-point, the Hamiltonian acting on a flavor $\alpha=1,2$ (or equivalently spin polarization) reads:
\begin{equation}
{\cal H}^{\pm}_{\alpha}({\bf k}) = \begin{pmatrix}
-d_{\alpha z}(K) & A_0 e^{\pm i\omega t}+v_F\Pi^* \\
v_F\Pi +A_0e^{\mp i\omega t} & +d_{\alpha z}(K) \\
 \end{pmatrix} \quad 
\end{equation}
with $\Pi=p_x+ip_y = |{\bf p}| e^{i\tilde{\varphi}_{\alpha}}=\mp|{\bf p}|e^{i\varphi}$. We use the precise correspondence introduced in the Letter page 2. Here, $\mp$ refers to the sphere $1$ and sphere $2$ respectively and we have redefined $d_{1z}=m_1$ and $d_{2z}=-m_2$ in accordance with the definitions of ${\bf d}_1$ and ${\bf d}_2$ in the article. Now, we proceed in a similar way as for Nuclear Magnetic Resonance.  To identify the rotating frame, we re-define  $|a'_{\alpha}\rangle =e^{\mp i\frac{\omega t}{2}} |a_{\alpha}\rangle$ and $|b'_{\alpha}\rangle =e^{\pm i\frac{\omega t}{2}} |b_{\alpha}\rangle$ associated to the rotation operator ${\cal U}(t)=e^{\mp i\frac{\omega t}{2}\sigma_z}$. In this rotated frame, the effective Hamiltonian close to the $K$ Dirac point takes the form
\begin{equation}
{{\cal H}^{\pm}_{eff,\alpha}}({\bf k}) = {\cal U}{\cal H}^{\pm}_{\alpha}({\bf k}){\cal U}^{-1}\pm \frac{\hbar\omega}{2}\sigma_z = \begin{pmatrix}
-d_{\alpha z}(K)\pm \frac{\hbar}{2}\omega & \alpha v_F|{\bf p}|e^{-i\varphi}e^{\mp i \omega t}+A_0 \\
\alpha v_F|{\bf p}|e^{i\varphi}e^{\pm i\omega t}+A_0 & +d_{\alpha z}(K)\mp\frac{\hbar}{2}\omega\\
 \end{pmatrix} \quad 
 \end{equation}
 where $\alpha=\mp$ in the matrix corresponds to $\alpha=1,2$, respectively. 

 \subsubsection{Resonance Situation for Sphere $1$}
  
  We assume $d_{1z}=m_1>0$ and $\omega>0$ such that at the $K$ point, only the right-handed light polarization resonates with the sphere $1$ through the equality $m_1=\frac{\hbar\omega}{2}$. We also fix the angle $\varphi=-\omega t$ to discuss small detuning effects from the Dirac points for sphere $1$. We study the lowest order response in $A_0$ and in $|{\bf p}|$. 
 
 Suppose we prepare the system in the lowest-energy band of the Haldane model in the ground state $|\psi_{1+}\rangle$ at time $t=0$ where we keep the specific
 form $|\psi_{1+}\rangle=\cos(\theta/2)|a'_1\rangle+\sin(\theta/2)|b'_1\rangle$ assuming $\theta\rightarrow 0$ close to the K-point. We allow smooth deviations of the angle $\theta$ from the Dirac point to emphasize
 that here we have a continuum of states. Then, solving the eigenstates in the rotating frame, we find the probability to be in the upper state $|b'_1\rangle$ at time $t$ 
\begin{equation}
\label{eq1}
|\langle b'_{1} |\psi_{1+}(t)\rangle|^2 = -{\cal A}'(\theta<\theta_c) \cos^2\left(\frac{(A_0-v_F|{\bf p}|)t}{\hbar}\right) + {\cal A}'(\theta>\theta_c) \sin^2\left(\frac{(A_0-v_F|{\bf p}|)t}{\hbar}\right).
\end{equation}
This quantity refers to the transition probability at the $K$-point and taking formally $\theta=0$ agrees with the formula for the nuclear magnetic resonance. On the other hand, keeping the forms of the smooth fields allows us to link with the topological properties as well.
Defining $N_+^1(t)=\langle \psi_{1+}(t) | \psi_{1+}(t)\rangle$, this implies
\begin{equation}
\label{eq2}
\frac{dN_+^1}{dt^2}= - \frac{C_1 (A_0-v_F|{\bf p}|)^2}{\hbar^2}.
\end{equation}

For the sphere $1$ at the north pole we have the identification ${\cal A}'(\theta>\theta_c)=C_1=+1$ if we move the boundary $\theta_c\rightarrow 0$. Precisely, at the $K$ Dirac point, this formula is in agreement with Eq. (\ref{density}).

Close to the $K'$ Dirac point, the left-handed light polarization now is at resonance for the same angle $\varphi=-\omega t$ and for the same light frequency $m_1=\hbar\omega/2$. If we evaluate $|\langle a_1'|\psi_{1+}(t)\rangle|^2$, we obtain a similar formula as Eq. (\ref{eq1})
with $\cos^2\frac{\theta}{2}\leftrightarrow \sin^2\frac{\theta}{2}$ since the inversion of the mass $d_{\alpha z}(K)=-d_{\alpha z}(K')$ is equivalent to modify the role of $|a'_1\rangle$ and $|b'_1\rangle$. 

 \subsubsection{Light Response from Sphere $2$}
 
 When the sphere $1$ is at resonance,  for states in the vicinity of the $K$ Dirac point such that $A_0\gg v_F|{\bf p}|$, then the sphere $2$ is described by the time-independent matrix
 \begin{equation}
{{\cal H}^{\pm}_{eff,2}}({\bf k})  = \begin{pmatrix}
-d_{2 z}(K)\pm m_1 & A_0 \\
A_0 & +d_{2 z}(K)\mp m_1\\
 \end{pmatrix} \quad.
 \end{equation}
 Below, we introduce the mass asymmetry $\delta m = (m_2-m_1)$. 
 
 At the $K$-point, the sphere $2$ dominantly couples to the $-$ light-polarization. Preparing the initial state as $|\psi_{2-}\rangle$ at time $t=0$ with $|\psi_{2-}\rangle=|b'_2\rangle$  and taking into
 account the time-evolution of this $2\times 2$ matrix gives the transition probability
 \begin{equation}
 |\langle a'_{2} |\psi_{2-}(t)\rangle|^2 = \frac{A_0^2}{\delta m^2 + A_0^2} \cos^2\frac{\theta}{2} \sin^2\left(\frac{\sqrt{\delta m^2 +A_0^2}t}{\hbar}\right).
 \end{equation}
 Now, for sphere $2$, we have the identification ${\cal A}'_{\varphi}(\theta>\theta_c)=-\cos^2\frac{\theta}{2}$ and also ${\cal A}'_{\varphi}(\theta>\theta_c)=C_2$ if we move $\theta_c\rightarrow 0$. 
 Including the contribution from the $K'$ point due to the $+$ light polarization, then the transition probabilities for sphere $2$ become symmetrically equal to
 \begin{equation}
 \label{P2}
|\langle a'_{2} |\psi_{2-}(t)\rangle|^2 = \frac{A_0^2}{\delta m^2 + A_0^2} (-C_2) \sin^2\left(\frac{\sqrt{\delta m^2 +A_0^2}t}{\hbar}\right).
 \end{equation}
 For the sphere $2$, due to the inversion between lowest and upper bands compared to sphere $1$, $C_1$ becomes $-C_2$ in that formula such that $-C_2>0$. We emphasize here that $|C_i|$ occurs in the inter-band transition probabilities in the rotating frame in agreement with the $C^2_i$ in the original frame and the photo-induced currents. Developing the formula at short times, we observe that the transition probabilities are independent of the mass asymmetry $\delta m$ traducing the robustness of the topological phase towards this perturbation.
 
For the Kane-Mele model, since we have defined ${\bf d}_1$ and ${\bf d}_2$ as opposite and radial vector fields on the sphere, this requires to define the polar angles around the $K$ Dirac point as $\tilde{\varphi}_1=\varphi\pm \pi$ and $\tilde{\varphi}_2=\varphi$ for each spin polarization. Since the physical energy spectrum is independent of $\varphi$, these choices are applicable. This implies that when we fix $\varphi=-\omega t$ for sphere $1$ at resonance then deviations from the poles on sphere $2$ lead to small corrections in the matrix $v_F|{\bf p}|e^{\pm 2i\omega t}$. 

\subsection{Interaction Effects and Spin Pump}

Here, we discuss the protection of the photo-induced response at the poles in the presence of interactions as long as we stay in the topological insulating phase and the relation with a $\mathbb{Z}_2$ spin pump. We assume here an interaction in real space which can also involve nearest-neighbors on different sublattices, and we study the most dominant
interaction channel(s) from the topological ground state.

We start from the ground state situation with the two lower energy bands occupied at half-filling (one related to each sphere). Locally, at the two Dirac points, we have two classes of interaction. At the $K$ point, the ground state satisfies the projection equalities $\hat{N}_1^a |GS\rangle=1$ and $\hat{N}_2^b |GS\rangle=1$ with  $\hat{N}_1^b |GS\rangle=0=\hat{N}_2^a |GS\rangle$. From the spin-1/2 quantum Lie algebra we have the operator related to the number of particles written in terms of the projectors 
\begin{equation}
\hat{N}_1^{i} = \frac{1}{2}\left(1\pm \sigma_{z}^1 \right), \hskip 1cm \hat{N}_2^{i} = \frac{1}{2}\left(1\pm \sigma_{z}^2 \right).
\end{equation}
Here, $i=a$ and $i=b$ refer to the polarisation (projection) on a given sublattice $A$ or $B$ of the honeycomb lattice. We have the identification $N^1(K) = \langle \hat{N}_1^a\rangle=1$ 
and similarly $N^1(K') = \langle \hat{N}_1^b\rangle=1$ for the lower band related to sphere $1$. The dominant Hubbard interaction at the $K$ Dirac point is of the form at low energy 
\begin{equation}
\label{interaction}
{\cal H}_{Int}=\lambda \left(\hat{N}_1^a \hat{N}_2^b\right).
\end{equation}
Similarly at the $K'$ point, the dominant interaction is between $\hat{N}_1^b \hat{N}_2^a$. The interaction terms at the $K$ and $K'$ points projected on the ground state then take the form 
\begin{equation}
{\cal H}_{Int} = -\frac{\lambda}{4}\sigma_z^1\sigma_z^2 +\frac{\lambda}{4}(1\pm\sigma_z^1 \mp\sigma_z^2).
\end{equation}
For the Kane-Mele model, this produces a ferromagnetic Ising interaction and Semenoff masses at the poles. 

We can include ${\cal H}_{Int}$ in the matrix representation at the poles of the two spheres and compare the energetics of different spin states $\{|a_1 a_2\rangle; |a_1 b_2\rangle; |b_1 b_2\rangle; |b_1 a_2\rangle \}$. At the north pole, then we verify that $|a_1 b_2\rangle$ remains the ground state as long as $\lambda<\hbox{min}(2m_1,2m_2)$ with $d_{1z}=m_1$ and $d_{2z}=m_2$. At the south pole, $|b_1 a_2\rangle$ also remains the ground state as long as $\lambda<\hbox{min}(2m_1,2m_2)$. Assuming that $\lambda$ satisfies this prerequisite, then we can argue the stability of the topological response from the correspondence
\begin{equation}
\label{Calpha}
C_{\alpha} = \frac{\langle\sigma_z^{\alpha}(0)\rangle-\langle\sigma_z^{\alpha}(\pi)\rangle}{2},
\end{equation}
and therefore of the light-response through the identification
\begin{equation}
(C_{\alpha})^2 = \frac{1}{4}\left(\langle\sigma_z^{\alpha}(0)\rangle^2 + \langle\sigma_z^{\alpha}(\pi)\rangle^2 -2\langle \sigma_z^{\alpha}(0)\rangle\langle\sigma_z^{\alpha}(\pi)\rangle\right) = \frac{1}{4v_F^2}\left({\cal I}_{\alpha}(0)+{\cal I}_{\alpha}(\pi)\right),
\end{equation}
with ${\cal I}_{\alpha}(\theta)$ as in Eq. (10) of the Letter for a sphere $\alpha$. This also implies that ${\cal A}_{\varphi}^{\alpha}(0)-{\cal A}_{\varphi}^{\alpha}(\pi)$ is unchanged and similarly for the Berry curvatures in Eq. (8) of the Article. As long as the structure of the lower and upper bands remains identical the light responses will keep a similar form if we shift the resonance frequency with $\lambda$. 

Since the light response at the poles corresponds effectively to measure $C_1-C_2$, either as $C_1^2+C_2^2$ in the original frame or as $C_1+|C_2|$ in the rotated frame, we emphasize here that this can also be re-interpreted as a spin pump measurement driving from north to south pole in the adiabatic limit.  Indeed, we may define 
\begin{equation}
\tau_z = \frac{1}{2}\left(\sigma_z^1 -\sigma_z^2\right)
\end{equation}
such that $\tau_z=\pm 1$ at the poles of the sphere corresponds to the spin magnetization if $\alpha=1,2$ refers to the spin polarization of an electron. In this way, we have
\begin{equation}
C_1 - C_2 = \langle \tau_z(0)\rangle - \langle \tau_z(\pi)\rangle = -\int_0^{\frac{\pi}{v}} \frac{\partial\langle \tau_z(t)\rangle}{\partial t} dt
\end{equation}
with the identification between polar angle on the sphere and time such that $\theta=vt$. As shown in the article, the spin pump can be activated through an electric field corresponding to a drive from north to south poles.
This makes a link between the responses at the poles of the sphere and an effective spin pump.

We can also include interaction effects between different Dirac points (if we also include the on-site Hubbard interaction in real space):
\begin{equation}
{\cal H}_{Int}' = \lambda' \left(\hat{N}_1^a({\bf K}) \hat{N}_2^a({\bf K}') + \hat{N}_1^b({\bf K}') \hat{N}_2^b({\bf K})\right).
\end{equation}
The energy $E_{a_1 b_2}+E_{b_1 a_2}$ associated to the states $|a_1 b_2\rangle$ at north pole and $|b_1 a_2\rangle$ at south pole will increase by $2\lambda'$ such that 
\begin{equation}
E_{a_1 b_2}+E_{b_1 a_2} = -2(m_1+m_2) +2\lambda+2\lambda' .
\end{equation}
The ground state at the two poles remains unchanged as long as $\frac{1}{2}(E_{a_1 b_2}+E_{b_1 a_2})<\{E_{a_1 a_2};E_{b_1 b_2}\}$ implying then $(\lambda+\lambda')<\hbox{min}(2m_1,2m_2)$. 
Therefore, we verify that the light response at the poles is protected (at least) as long as the effective interaction $\lambda+\lambda'$ is typically smaller than the energy band gap at the poles.

\end{document}